\documentclass[aps,prx,reprint,superscriptaddress,longbibliography,bibnotes]{revtex4-2}

\usepackage{graphicx}
\usepackage{amsmath,amssymb}
\usepackage{bm}
\usepackage{tikz}
\usetikzlibrary{arrows.meta,positioning}
\usepackage[colorlinks=true,linkcolor=blue,citecolor=blue,urlcolor=blue]{hyperref}

\colorlet{MAGENTA}{magenta}
\colorlet{BLUE}{blue}
\colorlet{RED}{red}

\begin{document}

\title{A fractional quantum Hall factory on quantum processors: constant-depth preparation of clustered non-Abelian states}

\author{Cheng Xu}
\affiliation{Department of Physics and Astronomy, University of Tennessee, Knoxville, TN 37996, USA}
\affiliation{Max Planck Institute for Chemical Physics of Solids, 01187 Dresden, Germany}

\author{Ching Hua Lee}
\affiliation{Department of Physics, National University of Singapore, Singapore}

\author{Hong-Hao Tu}
\affiliation{Faculty of Physics and Arnold Sommerfeld Center for Theoretical Physics, Ludwig-Maximilians-Universität München, 80333 Munich, Germany}

\author{Yang Zhang}
\email{yangzhang@utk.edu}
\affiliation{Department of Physics and Astronomy, University of Tennessee, Knoxville, TN 37996, USA}
\affiliation{Department of Physics, National University of Singapore, Singapore}

\date{\today}

\begin{abstract}
Non-Abelian anyons arise as exotic excitations in fractional quantum Hall (FQH) matter and have proved very elusive to realize in conventional platforms.
In this work, we show that on a programmable quantum hardware platform, the more exotic FQH excitations are the less costly ones to prepare: clustered non-Abelian FQH states admit parallel quantum preparation circuits whose two-qubit depth is independent of system size, while constructing the more common Abelian Laughlin state requires a sequential circuit chain with linear depth. 
The centerpiece of this work is our new systematic framework for cataloging possible FQH states and preparing them on quantum circuits at unprecedented scale and variety. Our prepared parafermionic Read--Rezayi $\mathbb{Z}_3$ state holds depth 3 from 8 to 118 qubits, and full root sampling extends to a 154-qubit, 104-electron Read--Rezayi $\mathbb{Z}_4$ state. In all, our demonstrated 18-family catalog of prepared FQH states extends to all 156 qubits of an IBM Heron processor, limited only by existing hardware scale.  
Measurements on the prepared states recover the expected fractional quasihole charges, with the charge estimator exact in every symmetry-selected shot for the clustered states, and braiding data of the non-Abelian $e/4$ quasihole measured via interferometric extensions. Our work establishes a scalable route to studying FQH physics on quantum processors and opens new avenues for preparing and probing non-Abelian topological matter far beyond the reach of conventional platforms.

\end{abstract}

\maketitle

\section{Introduction}
\begin{figure*}[t]
\includegraphics[width=\textwidth]{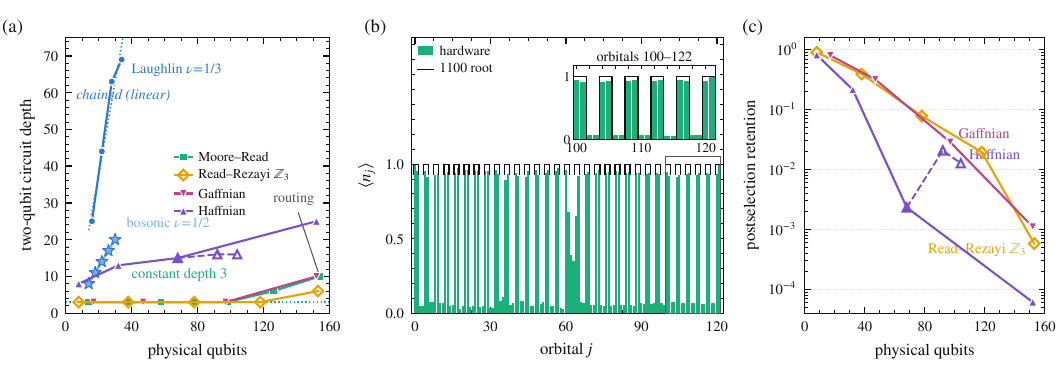}
\caption{The clustering dichotomy (linear vs. constant circuit depth) for state preparation across our FQH catalog. (a) Routed two-qubit depth after transpilation against qubit count for six families of FQH states we prepared: the two chained Laughlin recursions, the fermionic $\nu=1/3$ and the dual-rail bosonic $\nu=1/2$, both grow linearly, while the clustered Moore--Read (to 98 qubits), Read--Rezayi $\mathbb{Z}_3$ (to 118), and Gaffnian (to 97) hold constant depth three, and the Haffnian, whose squeeze runs two levels deep, sits at depth 8 to 16. Open dashed markers are the depth-16 \texttt{ibm\_kingston} registers that extend the Haffnian clean ladder to 104 qubits. Beyond roughly 120 qubits, heavy-hex routing raises the depth. (b) Postselected density of the 62-electron Moore--Read ground state on 122 qubits at algorithmic depth three (\texttt{ibm\_kingston}, 245 shots surviving $(N,K)$ postselection). The $1100$ charge-density-wave root (solid outline) is recovered at all 62 electrons across the full register. (c) Postselection retention against qubit count for the three clustered ladders of the same single-job \texttt{ibm\_fez} run. At 153 qubits, the Read--Rezayi root is still recovered at every electron. }
\label{fig:dichotomy}
\end{figure*}

Topological order expanded the classification of quantum matter beyond symmetry breaking, and fractional quantum Hall (FQH) states stand out as a particularly clean paradigm, whose model wavefunctions admit exact parent Hamiltonians~\cite{TsuiStormerGossard,Laughlin1983,Haldane1983,TrugmanKivelson}. They exhibit many of the hallmarks of topological order: fractionally charged quasiparticles, hidden string order in place of a local order parameter~\cite{GirvinMacDonald}, quantized geometric responses from FQH anisotropy~\cite{AvronSeilerZograf,ReadViscosity,yang2017generalized} and, in paired and clustered phases, non-Abelian exchange statistics~\cite{MooreRead,Wen1991,ReadRezayi}.

Yet, beyond the simplest Laughlin state, FQH states that exhibit more exotic quasiparticle statistics also require sophisticated multi-particle clustering properties that have proved elusive in most experimental quantum platforms. To realize them, digital quantum processors are ideal platforms: their versatility and programmability have recently enabled the simulation of a wide variety of condensed-matter phenomena~\cite{shen2026simulating,smith2019simulating,gou2020tunable,koh2022simulation,Kirmani2022,frey2022realization,chertkov2023characterizing,chen2023high,liu2024simulating,yang2023simulating,Iqbal2024,shen2025observation,koukoutsis2024quantum,koh2024realization,koh2025interacting,shen2025robust,zhang2025observation,shen2026observation,chen2026robust,jiang2026one}.
They offer capabilities beyond conventional condensed-matter experiments~\cite{von202040,lee2018floquet,kim2019even,leonard2023realization,wang2024realization,hu2025high,xu2025multiple,ahn2024nonabelian,reddy2024nonabelian,chen2025robust,wang2025higher}, such as full microscopic control, direct sampling of many-body configurations, and interferometric access to wavefunction phases.
Preparing FQH wavefunctions on quantum hardware requires translating their correlations into local gates, and the required circuit depth reflects the entanglement structure of the state. More broadly, the same logic applies to trial states with local positive parent Hamiltonians~\cite{Haldane1983,TrugmanKivelson,LeeLeinaas2004,SeidelFu2005,BergholtzKarlhede,LPT2015}: the parent supplies amplitudes, certificates, and controlled deformations. FQH states also provide a stringent testbed because fractional charge, hidden order, and non-Abelian statistics can all be encoded in exactly defined zero modes.

Quantum hardware demonstrations have so far concentrated on the Abelian $\nu=1/3$ Laughlin state, through linear-depth circuits and measurements of its topological responses~\cite{Rahmani2020,Kirmani2022,Kirmani2023,Kirmani2025,Shen2026}, and exact sparse-state constructions on the sphere~\cite{Wu2026}.
Non-Abelian anyons have been braided on processors only within exactly solvable lattice models whose braiding data is built in by construction~\cite{Andersen2023,Iqbal2024,Xu2024,Lo2026}.
Every demonstration so far has thus only targeted a single wavefunction (Table~S1 of the Appendix~\cite{SM}).
Two questions follow: what sets the circuit cost of an FQH wavefunction, and whether the full catalog of trial states can be prepared within a single framework.

In this work, we put forward a comprehensive FQH factory that generates a broad family of FQH states (our FQH catalog) at well-defined and surprisingly modest quantum hardware cost, and we extract the principles that set that cost. In hardware terms, some key achievements are: 18 trial-wavefunction families prepared and verified on 156-qubit processors, with up to 104 electrons per register; general classes of non-Abelian clustered states prepared at a circuit depth independent of system size; fractional charges counted shot by shot; and the braiding data of Moore--Read quasiholes measured interferometrically.

Our starting point is each family's pattern-of-zeros data, from which we construct and certify its defining parent Hamiltonian (Sec.~\ref{sec:parentH}). Within the pseudopotential parent-Hamiltonian~\cite{Haldane1983,TrugmanKivelson,LeeLeinaas2004,SeidelFu2005,BergholtzKarlhede,simon2007pseudopotentials,Lee2013,davenport2012multiparticle,lee2014lattice,LPT2015,seiringer2020emergence} catalog considered here, we include the single-component families and the multicomponent extensions whose channel projectors remain positive semidefinite and whose kernel counts match the pattern-of-zeros identities, the counting rules that encode each family's characteristic clustering and quasiparticles. These criteria are enforced as tests, and candidates that fail are excluded. For each of the 18 trial-wavefunction families studied here, at its natural filling fraction, we build a positive semidefinite Hamiltonian on the cylinder whose exact zero mode is the target state. 
We combine this catalog with a systematic thin-torus circuit construction and with symmetry-verified sampling, and we execute the resulting factory, a parent-Hamiltonian-to-circuit pipeline, on IBM Heron processors. 

The pseudopotential Hamiltonian makes a trial FQH wavefunction operational on hardware. 
It fixes the conserved quantum numbers used for error-heralding postselection and certifies the prepared state through zero-mode and counting checks.
The pinning fields (local potentials added to $H$ that trap quasiholes at chosen orbitals) that localize fractionalized domain walls and the metric and flux deformations behind the geometric-response protocols are deformations of the same Hamiltonian. Sec.~\ref{sec:whyH} spells out this logic in full. 
We obtain two structural results from this construction. 
We show that the local pair-hop amplitudes that squeeze the thin-torus Laughlin root patterns all obey a single closed form: a binomial coefficient multiplied by a circumference-dependent Gaussian [Eq.~\eqref{eq:universal}]. The known $\nu=1/3$ amplitude appears as the simplest special case.
Analogous closed forms hold across the whole catalog, including the Moore--Read block amplitude $\tau_{\rm MR} = -2\,e^{-2\kappa^2}$, so every preparation angle is analytic.

A key observation is that the circuit cost follows one of two scaling behaviors set by the root pattern: constant two-qubit depth when the root clusters, and linearly growing depth for the unclustered Laughlin series. We refer to this split as the clustering dichotomy, which has crucial implications for the scalability of FQH state preparation: as we shall see, more esoteric FQH states with more sophisticated cluster rules also have electrons kept further apart, and are hence surprisingly less costly to prepare fundamentally. 
Clustered states (Moore--Read, Read--Rezayi, Gaffnian, Halperin) squeeze only at junctions between charge clusters. The junction blocks share no electrons and execute in parallel at constant depth, the circuit expression of the clustered thin-torus vacuum structure identified for the Moore--Read state~\cite{BergholtzPfaffianTT} and extended to the full Read--Rezayi series, whose domain-wall counting matches the conformal-field-theory fusion rules~\cite{ArdonneTT}. 
The Laughlin series squeezes at every electron pair, adjacent blocks obstruct each other, and the circuit is the sequential chained recursion~\cite{Rahmani2020}.

With the state construction based entirely on the local clustering properties of the FQH quasiparticles, the circuit depth is independent of the system size, and we confirm this at scale on quantum hardware (Fig.~\ref{fig:dichotomy}).
Throughout, we distinguish the algorithmic two-qubit depth, its routed value after embedding in the device's degree-three heavy-hexagon (heavy-hex) lattice, and the total transpiled CZ count.
The parafermionic Read--Rezayi $\mathbb{Z}_3$ state holds an algorithmic two-qubit depth of three from 8 to 118 qubits, with the $11100$ root recovered at all 72 electrons and the same root still recovered at 93 electrons on 153 qubits. The non-Abelian Moore--Read, the Gaffnian, and the Haffnian ladders scale the same way (Sec.~\ref{sec:scale}).
Full root sampling extends to the 154-qubit, 104-electron Read--Rezayi $\mathbb{Z}_4$ state, and catalog circuits reach all 156 qubits of the device (Sec.~\ref{sec:tail}).
At the largest sizes, the dominant cost comes from heavy-hex routing, and postselection retention, not gate depth, becomes the limit. A first-principles gate-error model of the retention traces this limit to per-qubit readout rather than to the gate count, and reproduces the measured ladders (Sec.~S10 of the Appendix~\cite{SM}).

As a consequence, the non-Abelian clustered states, the non-unitary Gaffnian and Haffnian, and the multicomponent members are cheaper to prepare using our approach than the Abelian Laughlin series (Fig.~\ref{fig:factory}): their particles come in clusters of $k \geq 2$ ($k$ particles can meet, but not $k+1$ particles), so their junction squeezes stay disjoint. We prepare the Read--Rezayi $\mathbb{Z}_3$ parafermion state at a constant two-qubit depth of three, with three to nine transpiled CZs across the ladder and $93\%$ postselection retention, while the Laughlin circuit at the same electron number needs 42 CZs at linearly growing depth and retains $46\%$.

\begin{figure}[t]
\includegraphics[width=\columnwidth]{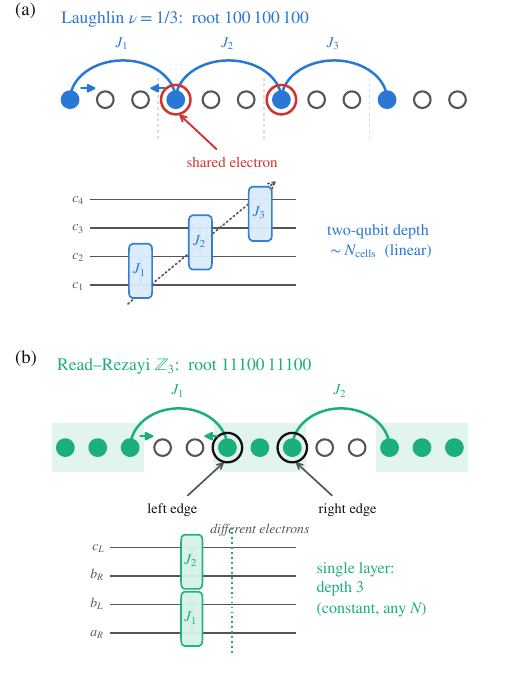}
\caption{
How the 2-qubit gate depth scales depends crucially on the clustering properties of the FQH state to be prepared. (a) Laughlin $\nu=1/3$ (root $100\,100\,100$): consecutive junction squeezes share an electron, so the flag rotations form a sequential chain and the depth grows with size. (b) Read--Rezayi $\mathbb{Z}_3$ (root $11100\,11100$): junctions couple disjoint electron pairs and fire in parallel at constant depth.
The junction squeezes $J_i$ and their amplitudes are defined in Sec.~\ref{sec:circuits} [Eq.~\eqref{eq:universal}], the flag-rotation circuits in the same subsection, and the four-family gallery with circuit sketches in Fig.~S2 of the Appendix~\cite{SM}. }
\label{fig:junctions}
\end{figure}

\emph{Clustering criterion for circuit scaling.} For the static thin-torus unitary circuits considered here, FQH state construction requires constant circuit depth when the allowed junction squeeze operations factor into disjoint cell-local blocks, as sketched in Fig.~\ref{fig:junctions}. When adjacent squeezes share a particle, the flag rotations form a blocking chain and the depth grows with the cell count. For the single-component pseudopotential families studied here, this occurs when every root cell carries a cluster of $k \geq 2$ particles and the parent adds no separate two-body channel. The multicomponent pair-channel parents factor conditionally, cell by cell (Sec.~S4 of the Appendix~\cite{SM}). 
The Laughlin series is the $k = 1$ boundary case, where cluster junction and electron pair coincide and the blocks fuse into a chain.
Interestingly, the junction count is blind to entanglement: the Halperin bilayer cell is more entangled than the Laughlin cell, yet it is prepared at constant depth (Sec.~S4 of the Appendix~\cite{SM}).

We demonstrate the Abelian-clustered corner on hardware as well, through the Halperin 331 family and the two-component $(3,3,2)$ and $(2,2,1)$ scaling ladders of Sec.~\ref{sec:tail}. The bosonic $(2,2,1)$ ladder holds two-qubit depth three out to 142 qubits, and the Halperin 331 root family runs on the full 156-qubit register.
The criterion is independent of the anyon statistics: the $\nu=1/4$ Pfaffian is non-Abelian yet compiles to the same sequential chain as the Laughlin series, because its parent needs the two-body $m=1$ channel alongside the three-body pseudopotentials, and that pair channel reintroduces squeezing at every electron pair.

A second, independent axis is the component number of the internal symmetry. The per-cell depth constant grows with it, by an order of magnitude from the SU(2) Bell cells (two-orbital entangled pairs) to the SU(3) GHZ cells (three-orbital entangled triplets; Sec.~\ref{sec:tail}), yet stays flat in system size, as the two three-component SU(3) ladders demonstrate to 141 qubits (Sec.~\ref{sec:tail}).

We recover the conclusion that the multicomponent structure of these Abelian members is necessary for $k\ge2$ clustering without non-Abelian statistics. For the single-component strong-pairing root $11000000$, an exact-diagonalization search over the admissible pseudopotential channels finds no positive parent with a clean zero-mode gap (Sec.~S2 of the Appendix~\cite{SM}). Within the pseudopotential channels~\cite{LPT2015} and system sizes searched here, a gapped single-component clustered parent appears only with non-Abelian statistics.

Constant depth is not excluded for the chained states in principle, since their thin-torus circuits are exact low-bond-dimension sequential matrix-product states, and measurement-assisted protocols for such states are a separate question beyond the scope of this work.

\begin{figure}[t]
\centering
\begin{tikzpicture}[node distance=2.6mm, box/.style={draw, rounded corners=1.5pt, align=flush left, inner sep=3.5pt, text width=0.88\columnwidth, font=\footnotesize}]
\node[box] (a) {\textbf{Catalog}: pseudopotential parent Hamiltonians $H=\sum_M B_M^\dagger B_M$ [Eq.~\eqref{eq:parent}] constructed for 18 trial-wavefunction families from their pattern-of-zeros data (Table~\ref{tab:factory})};
\node[box, below=of a] (b) {\textbf{Acceptance tests} (filters): positivity and symmetry, kernel count against the pattern-of-zeros identity, zero-mode uniqueness and thin-torus root collapse, convention checks};
\node[box, below=of b] (c) {\textbf{Thin-torus circuit}: root pattern plus closed-form squeeze angles [Eq.~\eqref{eq:universal}]; constant two-qubit depth for clustered roots, chained recursion for $k=1$};
\node[box, below=of c] (d) {\textbf{Hardware run}: heavy-hex transpilation, execution, and symmetry-selected sampling on the conserved $(N,K)$};
\node[box, below=of d] (e) {\textbf{Topological responses}: fractional charge $Q(j)$, string order, spectral flow, momentum polarization, spin shift, braiding (Table~\ref{tab:responses})};
\foreach \x/\y in {a/b, b/c, c/d, d/e} \draw[-{Stealth[length=1.6mm]}] (\x)--(\y);
\end{tikzpicture}
\caption{Workflow of our FQH state construction and measurement ($N$: particle number; $K$: total momentum around the cylinder; the kernel count is the zero-mode degeneracy of Eq.~\eqref{eq:parent}). Each family's pattern-of-zeros data enters at the top box. We construct its parent Hamiltonian from that data and certify it back against the same data (kernel count against the $(k,r)$-admissibility count, thin-torus limit against the root pattern), and only the certified zero-mode kernel then supplies the circuit amplitudes, the acceptance certificates, the postselection quantum numbers, and the response protocols.
}
\label{fig:workflow}
\end{figure}
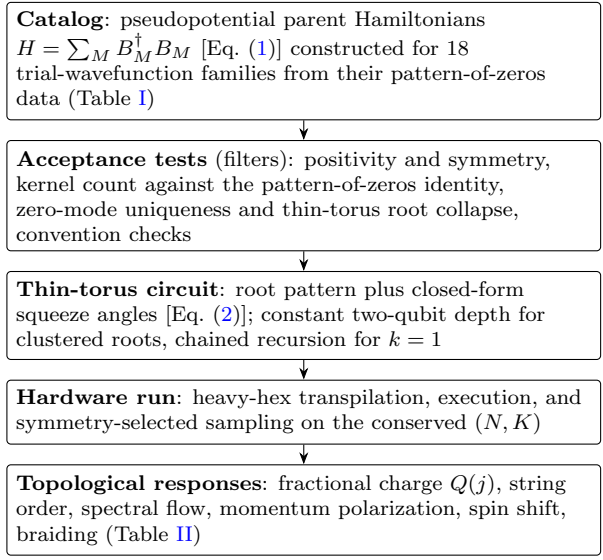

The paper follows the workflow (Fig.~\ref{fig:workflow}): the catalog and its circuits (Sec.~II), their scaling on hardware (Sec.~\ref{sec:scale}), the charge and hidden-order measurements (Sec.~III), the topological responses (Sec.~IV), and the route to the isotropic limit (Sec.~V).

\section{Models, circuits, and verification}

\subsection{Parent Hamiltonians on the cylinder}
\label{sec:parentH}

We work in the Landau-orbital basis of a finite cylinder with circumference $L_y$ (magnetic length set to one).
Orbital $j$ is a momentum eigenstate around the cylinder localized at guiding center $x_j = \kappa j$, with $\kappa = 2\pi/L_y$ the orbital spacing, and a many-body state is a superposition of occupation configurations of $N_{\mathrm{orb}}$ orbitals, one qubit per orbital under the Jordan--Wigner map for the single-component fermionic families (the bosonic and multicomponent encodings follow with the circuits). Near the thin-cylinder limit of vanishingly small $L_y$ (or large $\kappa$), the problem maps onto a one-dimensional lattice model~\cite{TaoThouless,SeidelFu2005,BergholtzKarlhede}. The isotropic regime is reached separately, through the bond-dimension construction of Sec.~\ref{sec:wall}. 

A family enters Table~\ref{tab:factory} only after its Hamiltonian is constructed and passes the acceptance tests of this section. We anchor each family to a pseudopotential parent Hamiltonian that we construct from its defining data, the pattern of zeros and root pattern taken from the trial-state literature, and check the integrity of our implementation (certify) back against that data. In the Landau-orbital basis, the pseudopotential Hamiltonians generically take the form~\cite{Haldane1983,TrugmanKivelson,LPT2015}
\begin{equation}
\begin{aligned}
H &= \sum_M B_M^\dagger B_M,\\
B_M &= \!\!\sum_{\substack{n_1<\cdots<n_p\\ \sum_i n_i = M}}\!\! P_\lambda(\bar n_1,\dots,\bar n_p)\, e^{-\frac{\kappa^2}{2}\sum_i \bar n_i^2}\, c_{n_1}\cdots c_{n_p},
\end{aligned}
\label{eq:parent}
\end{equation}
a sum over center-of-mass sectors $M$ of $p$-body cluster annihilation operators, with $\bar n_i = n_i - M/p$ the cluster-relative coordinates and $P_\lambda$ the polynomial that enforces the family's pattern of zeros, which encodes its defining clustering properties. The displayed single-component notation is shorthand, with bosonic and multicomponent parents using the corresponding mode- and component-resolved sums. The parent is frustration-free, so its zero modes are exactly the states annihilated by every $B_M$.

Based on the workflow in Fig.~\ref{fig:workflow}, we assemble the FQH channel content family by family. The channel polynomials $P_\lambda$ are literature input. We construct the second-quantized matrix of Eq.~\eqref{eq:parent} on the finite orbital register at each size and geometry, with the mode sums, Gaussian weights, and convention choices explicit. Our entire catalog spans (full channel specifications in Sec.~S2 of the Appendix~\cite{SM}): 
\begin{itemize}
\item the fermionic \textbf{Laughlin} series ($\nu = 1/q$, $p=2$): pair channels in the odd relative momenta $m \in \{1, 3, \dots, q-2\}$, whose $m=1$ member is the Haldane pseudopotential~\cite{Haldane1983,TrugmanKivelson}; root $100\cdots$, one particle per $q$-orbital cell;
\item the bosonic \textbf{Laughlin} state ($\nu = 1/2$, $p=2$): the $m=0$ contact channel, in the dual-rail encoding; root $1010\cdots$;
\item the paired \textbf{Moore--Read} state ($\nu=1/2$, $p=3$): the antisymmetric three-body Vandermonde channel~\cite{MooreRead,GreiterWenWilczek}; root $1100\,1100\cdots$;
\item the parafermionic \textbf{Read--Rezayi} $\mathbb{Z}_3$ and $\mathbb{Z}_4$ states ($p = 4, 5$): $k$-particle-cluster channels~\cite{ReadRezayi,simon2007pseudopotentials}; roots $11100\cdots$ and $111100\cdots$;
\item the non-unitary \textbf{Gaffnian} and \textbf{Haffnian} ($\nu = 2/5$, $1/3$): Moore--Read channels supplemented by higher-momentum triplet channels~\cite{SimonGaffnian,davenport2012multiparticle}, included as boundary stress tests;
\item the two-component \textbf{Halperin} bilayers 331, $(3,3,2)$, and $(2,2,1)$: intra-layer pair channels plus one or two inter-layer channels~\cite{Halperin1983,SeidelYang2008}; entangled layer-doublet roots;
\item the spinful-boson \textbf{NASS} and \textbf{spin-singlet Gaffnian} ($\nu = 4/3$, $4/5$): spin-symmetric contact channels~\cite{Ardonne1999} with $[H, \mathbf{S}^2] = 0$;
\item the three-component \textbf{SU(3)} states $(3,3,3\,|\,2,2,2)$ and $(2,2,2\,|\,1,1,1)$: color-symmetric channel sets with permutation-multiplet roots.
\end{itemize}
For the Laughlin series, the $B_M$ annihilate pairs of particles in the odd relative-momentum channels below $q$, and for Moore--Read they annihilate triplets through the antisymmetric $m=3$ polynomial~\cite{GreiterWenWilczek}. The Read--Rezayi $\mathbb{Z}_3$ state requires quadruplets, the Gaffnian adds the $m=5$ triplet channel~\cite{SimonGaffnian}, and the $\nu=1/4$ Pfaffian requires the two degenerate $m=9$ triplet pseudopotentials together with the two-body $m=1$ channel. The Haffnian combines the $m=3$, $5$, and $6$ triplet channels, the two-component Halperin 331 state combines intra-layer $m=1$ and inter-layer $m=0$ pair channels~\cite{Halperin1983}, and the spinful-boson non-Abelian spin-singlet (NASS) states~\cite{Ardonne1999} extend the same recipe with spin-resolved contact channels.
Every Hamiltonian conserves the particle number $N$ (per component) and the dipole $K = \sum_j j\, n_j$.

With the FQH channels decided on, every implemented candidate parent Hamiltonian (which need not coincide a priori with a particular pseudopotential) must pass the same four acceptance tests before any circuit work, and the tests filter candidates rather than admitting them by construction. 
The Hamiltonian must be positive semidefinite and conserve $(N, K)$ numerically.
Its kernel dimension must match the pattern-of-zeros count for the family, computed from the $(k,r)$ clustering rule alone~\cite{WenWang}, certifying that the constructed $H$ is the intended parent: the number of $(k,r)$-admissible occupation strings (at most $k$ particles in any $r$ consecutive orbitals) at general flux for the clustered series, and the documented excess or bound-saturation counts for the Haffnian and the spinful-boson sectors. We verified this identity at more than 70 flux and size points across the catalog, and it is the most discriminating of the acceptance tests (Sec.~S2 of the Appendix~\cite{SM}). The densest zero mode must be unique at the natural flux, and its thin-torus limit must collapse onto the family's root pattern.

The fourth test in Fig.~\ref{fig:workflow} fixes the residual ``convention'' choices through independent zero-energy benchmarks (details in Sec.~S2 of the Appendix~\cite{SM}), such as the Gaussian width of the channel weights at given $L_y$ and the pseudopotential admixture coefficients. 
Together, the tests in Fig.~\ref{fig:workflow} are able to rule out FQH scenarios that may a priori be assumed to exist: for instance, the single-component strong-pairing root $11000000$ admits no positive parent with a clean zero-mode gap in an explicit search of its channel space, while the strong-pairing phase is known to appear once a second component is available~\cite{ZhangRezayiYang2014}, the negative result of the Introduction that makes the multicomponent members necessary (Sec.~S2 of the Appendix~\cite{SM}).

\subsection{The operational role of the parent Hamiltonian}
\label{sec:whyH}

The trial FQH wavefunctions themselves have been known in closed form for decades, as first-quantized polynomials in the particle coordinates. From these trial states which identify the FQH families, our construction derives their certified parents, their closed-form amplitudes, and the catalog boundary.
A closed-form polynomial, however, is by itself neither a preparable object on hardware nor a checkable one. The circuit amplitudes, the certificates, and the response protocols all come from the parent Hamiltonian. 

First, the state amplitudes that a circuit needs to prepare are computed from the Hamiltonian.
A quantum register holds occupation configurations of Landau orbitals at one specific circumference $L_y$, so the preparation problem starts from the second-quantized amplitude list of the target state at that $L_y$, not from the polynomial. The circuit prepares this kernel state itself, which coincides with the trial wavefunction at the register's circumference. 
Converting the polynomial into that list is its own hard combinatorial problem. Jack-polynomial recursions solve it for the Laughlin and Read--Rezayi series~\cite{BernevigHaldane}, but no uniform tool covers the Gaffnian, the two-component states, or the spinful bosons, and kernel diagonalization fills that role.
We build $H$ in the fixed $(N, K)$ sector ($N$ the particle number, $K$ the total momentum around the cylinder), whose dimension stays polynomially small at thin-torus sizes. Its numerical kernel is the exact amplitude list of the target state at any $L_y$, for every family, by one procedure.

Every closed-form squeezing amplitude reported in this paper (for example the Laughlin $\tau_\ell$ of Eq.~\eqref{eq:universal} and the 16-channel NASS junction multiplet) was obtained by fitting the few circuit parameters to this kernel and recognizing the analytic form. After recognition, the analytic amplitudes are substituted back into the parent-Hamiltonian constraints and verified to annihilate the state at all tested sizes. The Moore--Read closed form $\tau_{\rm MR} = -2\,e^{-2\kappa^2}$ was recognized the same way, from the $N_e=4$ null space.
For the two-wall Read--Rezayi staircase of Sec.~\ref{sec:e5}, the symmetry-sector kernel is five-fold degenerate, so there the ``given wavefunction'' is not even unique, and the preparation target is well defined only as the circuit's projection onto the Hamiltonian's kernel.

Second, the parent Hamiltonian provides validation criteria for the target model and the ideal circuit, together with symmetry constraints for the measured hardware samples.
At 22 to 32 qubits, full state tomography cannot feasibly be performed, and a fidelity claim must rest on something the experiment can check. Concretely, the circuit acts on the root product state: at each junction a flag qubit is rotated by the closed-form angle, and a cascade of CX gates expands every flag into the squeezed configuration, cell by cell [Fig.~\ref{fig:junctions}(b)]. 
The parent Hamiltonian provides three such certificates.
We validate the model through the kernel-counting identity, the kernel dimension equaling the $(k, r)$-admissibility count. We then verify the built circuit state as a whole to be a zero mode, so the validation covers the compiled circuit rather than only the Hamiltonian model or its quoted kernel projection, at machine precision against exact diagonalization. 
During decoding, the conserved quantum numbers $(N, K)$ of $H$ are read off every shot. This symmetry verification heralds errors and, because the estimators of interest are quantized on the zero-mode manifold, upgrades ensemble averages into the per-shot counting statements of Secs.~\ref{sec:e4} and~\ref{sec:e5}.

Third, the observables that define the FQH phase are read from deformations of and additions to the Hamiltonian (flux threading, metric deformation, pinning potentials). 
Pinning potentials added to $H$ select the domain-wall and quasihole zero modes that carry the fractional charge ladder, flux threading generates the spectral flow and the pump staircase, and the metric deformation of the pseudopotentials generates the Hall-viscosity spin shift.
The quench experiment is the one exception: it evolves under $H$ itself. 
From the dipole bookkeeping of the $(N, K)$ sectors under wall transport we derive the interleaved position-parity ladders behind the braiding protocol, including the $e/4$ slope and the parity splitting the protocol measures.

Finally, our charge measurements are protected by an invariance of the Hamiltonian's zero-mode manifold: every configuration entering the prepared state carries the same fractional charge. 
Every squeezing channel of the zero mode moves zero charge across a cell boundary ($\Delta Q = 0$; Sec.~\ref{sec:e4}), while the $(N, K)$ sector as a whole admits dozens of values for the charge drop, and only $5\%$ of its configurations carry the target value.
The zero-mode manifold is therefore a strict subset of the $(N, K)$ sector, and the per-shot exactness of the $e/4$ and $e/5$ charges certifies that the kept shots satisfy the stronger constraint. 

\subsection{Constructing FQH ground states with squeezing operations}
\label{sec:circuits}

\begin{figure*}[t]
\includegraphics[width=\textwidth]{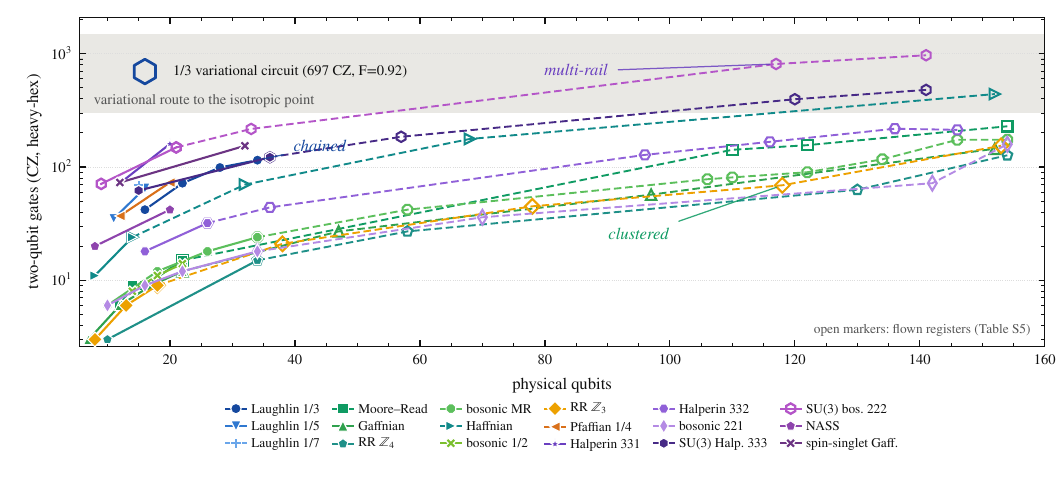}
\caption{Transpiled two-qubit cost against qubit count across the catalog. Filled markers are the exact-verification window, where every circuit carries an exact-diagonalization fidelity (Table~\ref{tab:factory}); open markers continue each family along its flown scaling ladder to chip scale, up to 154 qubits, at the routed depths listed in the footnotes of Table~S5 of the Appendix~\cite{SM}. Hue encodes the structural branch: chained Abelian (blues, shown over the verification window; their chip-scale depth scaling is in Fig.~\ref{fig:dichotomy}), clustered non-Abelian and non-unitary (greens and oranges), and multi-rail multicomponent (violets). The gray band is the variational route to the isotropic point (Sec.~\ref{sec:wall}).}
\label{fig:factory}
\end{figure*}

Near the thin-torus limit, each zero mode is generated from its root configuration by local squeezing operators $S_j = c^\dagger_{j+1} c^\dagger_{j+2} c_{j+3} c_j$ and their family-specific analogs, with amplitudes that we extract in closed form (Secs.~S3 and~S4 of the Appendix~\cite{SM}).
The Gaussian structure of thin-torus matrix elements has long been known~\cite{RezayiHaldane1994,SeidelFu2005,BergholtzKarlhede,LPT2015}. The kernel construction adds closed forms across the whole catalog, directly usable as circuit angles. For the Laughlin series, the amplitude to squeeze a root pair inward by $\ell$ orbitals is
\begin{equation}
\tau_\ell = (-1)^\ell \binom{q}{\ell}\, e^{-\kappa^2 \ell (q-\ell)}, \qquad \kappa = \frac{2\pi}{L_y},
\label{eq:universal}
\end{equation}
of which the known $\nu=1/3$ amplitude is the $q=3$, $\ell=1$ case.

The state $\prod_j (1 - t\, S_j)\lvert \mathrm{root}\rangle$, with $t \equiv -\tau_1 > 0$, maps onto a three-stage circuit: an X layer writes the root, rotations on flag qubits create the squeezing superposition, and a CX layer expands each flag into the squeezed configuration~\cite{Rahmani2020}.
The dichotomy of the Introduction enters at the flag stage.
For clustered states, the flags are independent and the stage is a single layer of $R_y$ rotations.
For the Laughlin series, a squeezed block forbids its neighbor from squeezing, and the flags form a chain of controlled rotations whose angles follow the recursion $\phi_{b-1} = \arctan(-t\cos\phi_b)$.
For $\nu=1/5$, each block carries a two-level rotation tree over the amplitudes $\tau_1, \tau_2$, and running the chain recursion with the combined amplitude $\lVert \tau \rVert_2$ reproduces the exact angles.
The Halperin 331 root is itself entangled, an equal-weight layer doublet in every charge cell, and its preparation adds one Bell-type block per cell plus a compensation layer that restores the equal-member measure distorted by the flag rotations.
Table~\ref{tab:factory} collects the transpiled costs, and Fig.~\ref{fig:factory} plots them across the catalog. Every circuit is verified against the exact zero mode by signed statevector overlap before submission, with fidelities between $0.96$ and $1.0$ across the verification window $L_y \lesssim 7$--$8$, wider than the window flown on hardware (Table~\ref{tab:factory}). 

This overlap is a dense-statevector check and so is confined to the exact-diagonalization window. Because the exact target and the constant-depth output are both small-bond-dimension matrix-product states, the same ideal-circuit fidelity is a transfer-matrix contraction computable at every flown size. It stays at or above $0.994$ at the largest registers tabulated in Table~S6 of the Appendix~\cite{SM} (Sec.~S3 of the Appendix~\cite{SM}).

\begin{table*}[t]
\caption{Our 18-family catalog realized on quantum hardware. ``Qubits/ED'' is the catalog verification window, and ``Qubits/Reach'' the largest hardware register flown, every entry a fully clean register. Two-qubit gate counts are transpiled heavy-hex CZ counts over the verification window. Fidelities are signed overlaps with the exact parent-Hamiltonian zero modes over the thin-torus window used on hardware. Retention is the fraction of shots surviving symmetry postselection, spanning each family's flown ladder. $F$ (MPS) is the ideal-circuit fidelity against the exact matrix-product target at the largest flown size. The upper block lists the chained families (linearly growing static-unitary depth) and the bold lower block the clustered families (constant depth). The fidelity methodology, the entangled-cell conventions, and the per-family flown provenance are in Sec.~S3 and Table~S6 of the Appendix~\cite{SM}. The most demanding entries are the 154-qubit Read--Rezayi $\mathbb{Z}_4$ and Moore--Read registers.
}
\label{tab:factory}
\begin{ruledtabular}
\footnotesize
\begin{tabular}{lcccccc}
Family & Qubits/ED & Qubits/Reach & CZ (ED) & $F$ vs ED & $F$ (MPS) & Retention \\
\hline
Laughlin $\nu=1/3$ & 16--34 & 34 & 42--115 & $\geq 0.996$ & -- & 0.02--0.60 \\
Laughlin $\nu=1/5$ & 11--16 & 16 & 35--65 & $\geq 0.996$ & -- & 0.43--0.59 \\
Laughlin $\nu=1/7$ & 15 & 15 & 69 & $0.998$ & -- & 0.40 \\
Bosonic $\nu=1/2$ & 14--22 & 22 & 8--14 & $\geq 0.998$ & -- & 0.73--0.82 \\
$\nu=1/4$ Pfaffian & 12--20 & 20 & 37--73 & $\geq 0.9999$ & -- & 0.48--0.72 \\
\hline
\textbf{Moore--Read $\nu=1/2$} & 10--22 & 154$^{\mathrm{a}}$ & 6--15 & $\geq 0.992$ & 0.9967 & 0.005--0.87 \\
\textbf{Read--Rezayi $\mathbb{Z}_3$ $\nu=3/5$} & 8--18 & 153 & 3--9 & $\geq 0.9986$ & 0.9993 & 0.003--0.93 \\
\textbf{Gaffnian $\nu=2/5$} & 7--22 & 152 & 3--12 & $\geq 0.988$ & 0.9942 & 0.008--0.82 \\
\textbf{Halperin 331 $\nu=1/2$} & 12--20 & 156 & 72--160 & $\geq 0.993$ & 0.9999 & 0.0085--0.59 \\
\textbf{Haffnian $\nu=1/3$} & 8--14 & 152 & 11--24 & $\geq 0.9999$ & 1.0000 & 0.00036--0.82 \\
\textbf{Read--Rezayi $\mathbb{Z}_4$ $\nu=2/3$} & 10--22 & 154 & 3--9 & $\geq 0.9993$ & 0.9997 & 0.0011--0.90 \\
\textbf{Bosonic Moore--Read $\nu=1$} & 10--14 & 154 & 6--9 & $\geq 0.999$ & 0.9940 & 0.0012--0.89 \\
\textbf{NASS $\nu=4/3$} & 8--20 & 92 & 20--42 & $\geq 0.9999$ & 0.9994 & 0.036--0.79 \\
\textbf{Spin-singlet Gaffnian $\nu=4/5$} & 12--32 & 32 & 73--154 & $\geq 0.9998$ & 0.9998 & 0.16--0.47 \\
\textbf{Halperin 332 $\nu=2/5$} & 16--26 & 146 & 18--32 & $\geq 0.9999$ & 0.9999 & 0.03--0.75 \\
\textbf{Bosonic Halperin 221 $\nu=2/3$} & 10--16 & 154 & 6--9 & $\geq 0.9999$ & 0.9999 & 0.0026--0.87 \\
\textbf{SU(3) Halperin 333 $\nu=3/7$} & 15--36 & 141 & 62--122 & $\geq 0.998$ & 0.9978 & 0.0023--0.60 \\
\textbf{SU(3) bosonic 222 $\nu=3/4$} & 9--21 & 141 & 71--149 & $\geq 0.9999$ & 1.0000 & 0.000023--0.60 \\
\end{tabular}
\end{ruledtabular}
\begin{minipage}{\textwidth}\footnotesize\setlength{\leftskip}{0pt}\setlength{\rightskip}{0pt}\setlength{\parfillskip}{0pt plus 1fil}
$^{\mathrm{a}}$ The Moore--Read register is recovered clean to 154 qubits, all 78 electrons, on \texttt{ibm\_boston}: excluding the two worst readout qubits from the coupling map and choosing a layout that keeps every occupied orbital on a clean qubit resolves every electron on 1\,309 postselected shots, with on-root density 0.945 and off-root density at most 0.100. The same broken-readout screening earlier completed the 150-qubit register on \texttt{ibm\_aachen}, all 76 electrons. Previous 150-qubit passes had placed orbitals 121 and 122 on qubits whose assignment error reached $0.41$, 75 times the device median. Removing them and the three next-worst readout qubits from the coupling map and rerouting the identical circuit resolved the final electron.
\end{minipage}
\end{table*}

\subsection{Symmetry-verified sampling}

All static observables in this work are diagonal in the occupation basis.
A hardware run therefore consists of preparing a circuit, measuring all qubits, and postselecting the bitstrings on the exact conserved quantum numbers of the parent Hamiltonian, and we call the surviving shots symmetry-selected. The conserved numbers are $N$ and $K$ for the single-component states, and the component-resolved particle numbers together with $K$ for the multicomponent and spinful states. The dual-rail encodings add an encoding-leakage veto, since each orbital mode occupies two qubits and any single bit flip changes a conserved quantity~\cite{Shen2026}.
Retention is a verification yield rather than a fidelity: every kept shot passes each symmetry check exactly, and all quoted statistics use kept shots alone. Because the symmetry filter is extensive, the retention decreases rapidly with register size even when the retained distribution remains accurate, so a low retention raises the sampling cost. Readout errors dominate that cost, removing about 27 times more shots per qubit than the gate errors (Sec.~S10 of the Appendix~\cite{SM}). 
The static results rely on this symmetry selection. Readout mitigation, applied where stated, shifts the postselected observables only at the $10^{-3}$ (0.1\%) level (the decoding and error-budget methodology is detailed in Secs.~S5 and~S7 of the Appendix~\cite{SM}).

All hardware data in this work were taken on the 156-qubit IBM Heron processors \texttt{ibm\_fez}, \texttt{ibm\_kingston}, \texttt{ibm\_marrakesh}, \texttt{ibm\_aachen}, and \texttt{ibm\_boston}, with gate and measurement twirling and XY4 dynamical decoupling, at 8192 to 16384 shots per circuit and 32768 shots for the largest scaling ladders.
Transpiled chip layouts at the flown maxima, with the broken readout qubits marked, are drawn in Fig.~S4 of the Appendix~\cite{SM}. %

The central response measurements replicate across \texttt{ibm\_fez} and \texttt{ibm\_kingston}: the $-e/3$ charge agrees between the two to better than $0.009$, and the spectral-flow and braid flux slopes agree within their quoted errors.
The chip-scale cells likewise replicate on \texttt{ibm\_fez}: the 154-qubit Read--Rezayi $\mathbb{Z}_4$ register recovers 103 of its 104 root peaks, while the 153-qubit $\mathbb{Z}_3$ and 156-qubit Halperin 331 registers recover every occupied root peak, on 87, 79, and 336 postselected shots respectively. The record claim rests on the pooled \texttt{ibm\_kingston} run of Sec.~\ref{sec:tail}, with the fez pass as the cross-device check.

\subsection{FQH state preparation with constant circuit depth}
\label{sec:scale}
Here, we elaborate on the quantum hardware verifications of the FQH states prepared at constant circuit depth. Due to the lack of cost-scaling, we achieved their preparation at the full scale of the device (Fig.~\ref{fig:dichotomy}): registers up to 156 qubits and 104 electrons run at two-qubit depths of 2 to 16. The ideal-circuit fidelity is validated by exact overlaps in the diagonalization window and by the matrix-product-state contractions of Table~\ref{tab:factory} at larger sizes. On quantum hardware, root coverage and postselection retention characterize the measured occupation distribution.

The non-Abelian Moore--Read ground state compiles to an algorithmic two-qubit depth of 3 at every size.
On a single \texttt{ibm\_fez} job, the depth stays at three out to 50 electrons on 98 qubits, with the $1100$ charge-density-wave pattern recovered at every electron [Fig.~\ref{fig:dichotomy}(a)].
Per-register transpilation extends the clean ladder to 64 electrons on 126 qubits, with all 64 peaks recovered on \texttt{ibm\_kingston} [Fig.~\ref{fig:dichotomy}(b) shows the 122-qubit register]. Beyond that size, the squeeze stars [Fig.~S4 of the Appendix~\cite{SM}] exhaust the lattice's degree-three vertices, as quantified below.
The parafermionic Read--Rezayi $\mathbb{Z}_3$ state runs at the same depth of three out to 72 electrons on 118 qubits, where the 637 postselected shots recover the $11100$ root at all 72 electrons.
The same root is recovered at 93 electrons on 153 qubits at a routing-inflated depth of six [Fig.~\ref{fig:dichotomy}(c)].
An \texttt{ibm\_kingston} replication reproduces the 118-qubit register at higher retention on 4\,352 postselected shots ($13\%$ retention), and shots pooled over seven preparations render the 153-qubit register clean, recovering all 93 electrons on 631 postselected shots (Fig.~S3 of the Appendix~\cite{SM}).
The Gaffnian matches its root comb at depth three to 97 qubits with all 40 electrons recovered. The Haffnian, whose squeeze runs two levels deep, compiles to a deeper but still size-modest two-qubit depth between 8 and 16, with all electrons recovered through 122 qubits.

Practically, cost is incurred by the factors: shot pooling, screened relayout, and per-register retranspilation. At full chip, the limiting resource however is retention, as evident in our \texttt{ibm\_kingston} implementation: the Gaffnian extends to 152 qubits, all 62 electrons recovered on 956 postselected shots pooled across passes [Fig.~S5(b) of the Appendix~\cite{SM}], and the 130-qubit Moore--Read root is recovered in full, 66 of 66 electrons on 139 shots pooled over two passes. Below, we provide the detailed account, with comparisons between different quantum processors.

A device-swap pass on \texttt{ibm\_aachen} completes the 152-qubit Haffnian root, all 52 electrons recovered on 70 postselected shots pooled across six passes. The same swap completes the 117-qubit SU(3) bosonic $(2,2,2)$ register, all 30 per-color root peaks on 245 postselected shots, where two \texttt{ibm\_kingston} passes had kept 18 and 16. The 150-qubit Moore--Read register completes clean on the same device once the two broken readout qubits carrying orbitals 121 and 122, and the three next-worst screened alongside, are excluded from the layout, all 76 electrons recovered on 2\,706 postselected shots, at roughly eightfold higher retention. Earlier passes had reached 75 of 76 because those orbitals sat on qubits whose assignment error reached $0.41$, a misassignment rather than a comb-level fidelity floor, and the flown job's own calibration data localize the defect (mechanics in Sec.~S5 and layouts in Fig.~S4 of the Appendix~\cite{SM}). The same broken-readout class is present on \texttt{ibm\_kingston}, so unscreened chip-scale layouts ride defective qubits on either device, and excluding them from the layout is the demonstrated cure.
The Haffnian root comb is recovered at all electrons through 104 qubits on \texttt{ibm\_kingston}, 32 of 32 and 36 of 36 electrons on 1\,365 and 830 postselected shots pooled across two passes at depth 15 to 16. The 122-qubit register completes clean on a screened \texttt{ibm\_aachen} relayout, all 42 electrons recovered on 2\,685 postselected shots at roughly 19-fold higher retention, the same broken-readout cure (Sec.~S5 of the Appendix~\cite{SM}).

Transpiling every register from scratch makes each family's chip-scale limit a measured quantity: retention and readout quality set it, and both are extended by the screened relayouts just shown. 
The Halperin 331 root-family circuit, the Bell-cell root with the exponentially small junction squeeze dropped (Table~\ref{tab:factory}), holds two-qubit depth three from 44 to 156 qubits and recovers every occupied root peak on all seven registers. Retention runs from $59\%$ down to $0.85\%$ at 156 qubits, where two chip-scale passes recover 382 postselected shots in all, clean end-to-end.
The bosonic Moore--Read dual rail stays clean at depth two to 122 qubits and carries to 154 qubits at routed depth nine, with all 39 root orbitals resolved on 200 postselected shots (Sec.~\ref{sec:tail}).
The fermionic Moore--Read instead meets a two-tier barrier.
Its junction squeeze is a three-leaf star that fills a single degree-three heavy-hex vertex, and the chip's 48 degree-three vertices pack only about 25 disjoint stars, so the zero-routing depth-three window saturates near 98 qubits.
Routed registers at depth six then carry the clean ladder to 130 qubits, and a broken-readout-screened relayout on \texttt{ibm\_boston} recovers the 78-electron register clean at 154 qubits (Table~\ref{tab:factory}, note~a).
Which families reach the full chip is therefore set by how the root cell tiles the heavy-hex lattice [Fig.~S4 of the Appendix~\cite{SM}], with the single-orbital and dual-rail clusters carrying to within a few qubits of 156 while the four-qubit-per-orbital NASS encoding reaches 92.

\section{Measurement of fractional charge and hidden order}
\begin{table*}[t]
\caption{Summary of the measured topological responses. Each response is read from symmetry-selected shots of the named state. The listed sections and figures describe the registers, devices, shot counts, and error budgets, and the last column names the dominant systematic error. }
\label{tab:responses}
\begin{ruledtabular}
\footnotesize
\begin{tabular}{p{2.7cm}p{2.5cm}p{3.6cm}p{1.4cm}p{3.6cm}p{1.6cm}}
Response (state) & Estimator & Measured & Exact & Main systematic error & Where \\
\hline
Charge $-e/3$ (Laughlin) & window charge $\sum_{j\in W}(\langle n_j\rangle-1/3)$ & $-0.3337(3)$ / $-0.3333(0)$ (two devices) & $-1/3$ & off-quantized zero-mode weight ($2.6\times10^{-4}$) & Sec.~\ref{sec:laughlin} \\
Charge $-e/4$ (Moore--Read) & $\Delta Q$ across pinned wall & $-1/4$ in all $20\,630$ shots & $-1/4$ & manifold-breaking weight $<1.5\times10^{-4}$ & Sec.~\ref{sec:e4}, Fig.~\ref{fig:mre4} \\
Charge $-e/5$ (RR $\mathbb{Z}_3$) & $\Delta Q$ across pinned wall & $-1/5$ in all $21\,105$ shots & $-1/5$ & manifold-breaking weight $<1.5\times10^{-4}$ & Sec.~\ref{sec:e5}, Fig.~\ref{fig:e5} \\
String order (Laughlin) & $O_{\mathrm{str}}$ [Eq.~\eqref{eq:obs}] & $0.04\to0.49$ across $L_y=5$--$10$, within $0.001$--$0.01$ of circuit-exact for $L_y\ge7$ & exact curve & representative bias at large $L_y$ & Sec.~\ref{sec:laughlin}, Fig.~\ref{fig:campaign1} \\
Spectral flow (Laughlin pump) & charge-transfer slope per $\Phi_0$ & $+0.3549(11)$ / $-0.3598(10)$ & $\pm0.36$ & in-sector noise shifts the crossing & Sec.~IV \\
Momentum polarization (Laughlin) & $\lambda(L_y)$ & tracks exact at three sizes, two devices & exact curve & central-charge offset inaccessible & Sec.~IV \\
Spin shift (Moore--Read) & finite-cylinder $S$ & within $1\sigma$ at all four $L_y$ & truncated-model curve & interferometric damping & Sec.~IV \\
Braid flux slope (MR $e/4$) & interference phase per $\Phi_0$ & $0.2500(17)$ / $0.2503(12)$ & $1/4$ & contrast damping, recalibrated & Sec.~IV \\
Fusion splitting (MR $e/4$) & loop-reversal phase gap & $3.10(3)$ / $3.02(3)$ & $\pi$ & decode drift, controlled by width scan & Sec.~IV \\
\end{tabular}
\end{ruledtabular}
\end{table*}

\begin{figure}[t]
\includegraphics[width=\columnwidth]{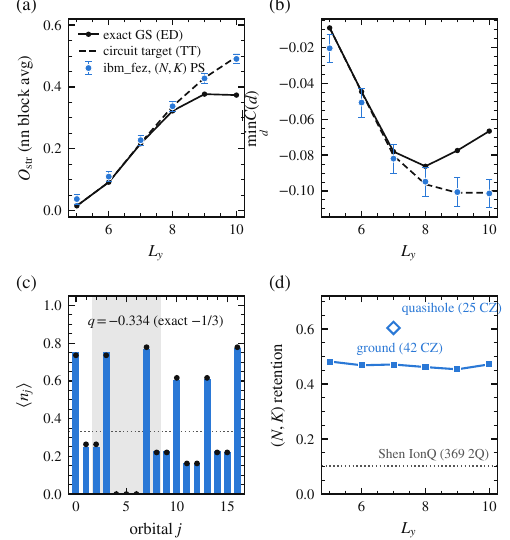}
\caption{Fixed-depth $L_y$ sweep of the $\nu=1/3$ Laughlin state on \texttt{ibm\_fez}, 8192 shots per circuit, readout-mitigated and $(N,K)$-postselected. (a) String-order crossover. (b) Correlation-hole formation. (c) Quasihole domain wall and window charge. (d) Postselection retention. Error bars are binomial shot noise $1/\sqrt{N_{\mathrm{kept}}}$; bootstrap intervals for the nonlinear estimators are given in Sec.~S7 of the Appendix~\cite{SM}. The \texttt{ibm\_kingston} replication agrees to $\pm 0.009$ throughout.}
\label{fig:campaign1}
\end{figure}

Having prepared our whole catalog of FQH states at scale, we proceed to measure smoking-gun order parameters for FQH states. We focus on the following two diagonal estimators
\begin{equation}
\begin{gathered}
Q(j)=\sum_{i\le j}\big(\langle n_i\rangle-\bar\nu\big),\\
O_{\rm str}(i,j) = -\Big\langle\, S^z_i \prod_{b=i+1}^{j-1}(-1)^{\,n_{3b+1}+n_{3b-1}}\, S^z_j \,\Big\rangle,
\end{gathered}
\label{eq:obs}
\end{equation}
the cumulative charge and the hidden string order, with $S_b^z=n_{3b+1}-n_{3b-1}$ for the Laughlin root cells. Table~\ref{tab:responses} maps every measured response to its estimator, result, and dominant systematic. 

\subsection{Laughlin quasiholes and the string-order crossover}
\label{sec:crossover}
\label{sec:laughlin}

Figure~\ref{fig:campaign1} shows the fixed-depth $L_y$ sweep of the $\nu=1/3$ state in the thin-torus limit, exactly prepared by our quantum circuit. 
The string order $O_{\rm str}$ [Eq.~\eqref{eq:obs}] detects the diluted antiferromagnetic pattern hidden in the root~\cite{GirvinMacDonald} and is read directly from the same occupation samples as the charge. Because the thin-torus circuit cost is independent of $L_y$, the same 42-CZ circuit family scans the crossover from the charge-density-wave limit to the correlated regime. 
In (a), the measured string order $O_{\rm str}$ [Eq.~\eqref{eq:obs}, summed over the central window] rises from $0.04$ at $L_y=5$ to $0.49$ at $L_y=10$ and closely follows the circuit-exact values to between $0.001$ and $0.01$ for $L_y \geq 7$. The site-averaged correlation hole deepens to $-0.101$ over the same sweep. 
The deviation of the prepared state from the true ground state beyond $L_y \approx 8$ is by design: it locates the breakdown of the thin-torus representative, the adiabatically connected small-circumference member of the same topological phase, which our circuits prepare exactly.

A domain-wall circuit at 25 CZs realizes the $e/3$ quasihole, and the window-summed excess charge reads $-0.3337(3)$ on \texttt{ibm\_fez} and $-0.3333(0)$ on \texttt{ibm\_kingston}, against the exact $-1/3$.
The estimator is itself quantized in multiples of $e/3$ on the prepared manifold (Sec.~S6 of the Appendix~\cite{SM}): every one of the 4\,839 symmetry-selected \texttt{ibm\_kingston} shots returned exactly $-1/3$, and 2 of 4\,950 deviated on \texttt{ibm\_fez}, consistent with the off-quantized estimator weight $2.6\times 10^{-4}$ the exact zero mode itself carries. 

Circumference and electron number play distinct scaling roles: $L_y$ drives the crossover physics at fixed circuit cost, while $N_e$ scales the register while keeping the physics fixed. 
At $N_e = 10$, the same construction reaches 28 qubits and extends the string-order plateau to block separation seven at $+0.148$. 
At $N_e = 12$, it reaches 34 qubits and recovers the $\nu=1/3$ charge-density-wave pattern at all 12 density peaks at $2\%$ retention, about 160 kept shots [Fig.~\ref{fig:campaign1}(d)]. 

\begin{figure}[t]
\includegraphics[width=\columnwidth]{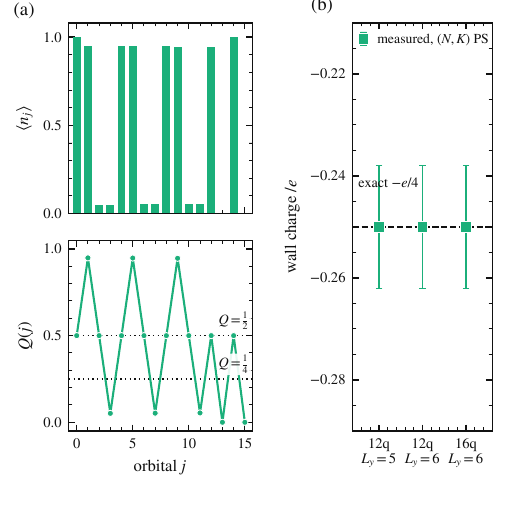}
\caption{Measured Moore--Read domain walls on \texttt{ibm\_fez}, 8192 shots per circuit, $(N,K)$-postselected at retention 0.79--0.86. (a) Measured density profile and cumulative charge $Q(j)$ of the 16-qubit wall. The cell-averaged step is the $e/4$ charge. The ideal profile is the paired $1100$ vacuum left of the pinned wall and the $1010$ vacuum to the right of it. The register terminates in the $1010$ pattern at the right edge. (b) The wall charge on all three circuits, read by the region-averaged step estimator (Sec.~S6 of the Appendix~\cite{SM}), whose binomial shot noise $1/\sqrt{N_{\mathrm{kept}}}$ sets the error bars. The per-shot cell-to-cell drops of the counting statement are exact and carry no spread. }
\label{fig:mre4}
\end{figure}

\begin{figure}[t]
\includegraphics[width=\columnwidth]{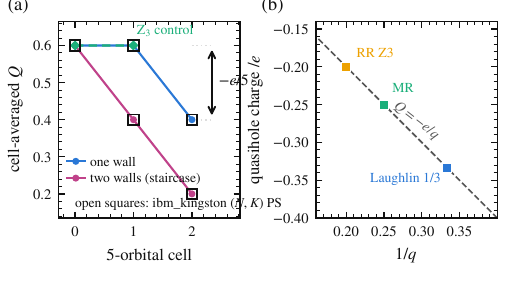}
\caption{The measured $e/5$ charge of the Read--Rezayi state on \texttt{ibm\_kingston}, 8192 shots per circuit, readout-mitigated and $(N,K)$-postselected. (a) Cell-averaged cumulative charge of the $\mathbb{Z}_3$ Read--Rezayi ground state, single wall, and two-wall staircase at $L_y = 5$ (13 to 15 orbitals, one qubit per orbital). Every postselected shot carries the quantized $-e/5$ drops. (b) The measured fractional charge ladder across the three host states on the two devices (\texttt{ibm\_fez} and \texttt{ibm\_kingston}). Error bars are smaller than the markers.}
\label{fig:e5}
\end{figure}

\subsection{The Moore--Read $e/4$ charge and its protection}
\label{sec:e4}

The Moore--Read state is the leading candidate for the observed $\nu=5/2$ plateau, so its $e/4$ quasiholes are the natural target for a braiding measurement.
The Moore--Read walls separate a paired $1100$ region from a $1010$ region, the two inequivalent thin-torus vacua whose domain walls carry $\pm e/4$~\cite{BergholtzPfaffianTT}, and both regions sit at $\nu = 1/2$, so no local density deficit marks the wall, and the charge appears only as a shift of the cumulative polarization.
We measure the cell-averaged cumulative charge $Q(j) = \sum_{i \leq j} (\langle n_i \rangle - \tfrac12)$, which steps from $\tfrac12$ deep in the paired region to $\tfrac14$ in the unpaired region.
On hardware, the result is a per-shot counting statement: every one of the 20\,630 symmetry-selected shots across the three wall circuits (12 and 16 qubits) carries exactly the cell-to-cell drop $\Delta Q = -1/4$, the estimator being quantized in quarter units (Fig.~\ref{fig:mre4}).
This bounds the weight of manifold-breaking configurations below $1.5\times 10^{-4}$ at $95\%$ confidence.

$Q$ is not an invariant of the $(N, K)$ symmetry sector, whose configurations span $Q \in [-1.25, 0.75]$. Every squeezing channel of the zero mode, however, carries $\Delta Q = 0$ exactly, so the wall charge is invariant under the state's own squeezing fluctuations, and the surviving noise averages out under twirling.
The Laughlin window charge lacks the exact version of this invariance: its squeezing channels can carry charge through the window boundary, so the zero mode itself holds a small off-quantized weight, $2.6\times 10^{-4}$, and the measured deviations sit at that level rather than at zero.
The contrast between exact and merely sharp quantization is set by whether the parent clustering confines squeezing within cells.

\subsection{The Read--Rezayi $e/5$ charge and the two-wall staircase}
\label{sec:e5}

The $\mathbb{Z}_3$ Read--Rezayi walls repeat the construction one level up. A $11100$ pattern meets a $11010$ pattern at $\nu = 3/5$, and the cell-averaged charge steps by $-e/5$, the domain-wall structure of the Read--Rezayi thin-torus vacua~\cite{ArdonneTT}.
Five circuits on \texttt{ibm\_kingston} at 13 to 15 qubits and at most 25 CZs prepared the ground state as a flat control, a single wall, and a two-wall staircase.
The staircase root's symmetry-sector kernel is five-fold degenerate, so we verified its circuit by projection onto the kernel, at weight $0.999$, rather than by overlap with any single member.
Postselection retained $92$ to $93\%$ of the mitigated weight, the highest retention in this work.
The result repeats the Moore--Read counting statement at the next denominator: all 21\,105 raw symmetry-selected shots carry exactly the quantized drops, $(0, -\tfrac15)$ across the single wall and $(-\tfrac15, -\tfrac15)$ down the staircase, bounding manifold-breaking weight below $1.5\times 10^{-4}$ at $95\%$ confidence.
Again the result goes beyond the symmetry sector alone. The $(N, K)$ sectors admit 45 to 49 distinct drop values, and only $5\%$ of sector configurations carry the target one. 
The measured ladder now spans $-e/3$, $-e/4$, and $-e/5$ (Fig.~\ref{fig:e5}), with both clustered charges per-shot exact.

\begin{figure*}[t]
\includegraphics[width=\textwidth]{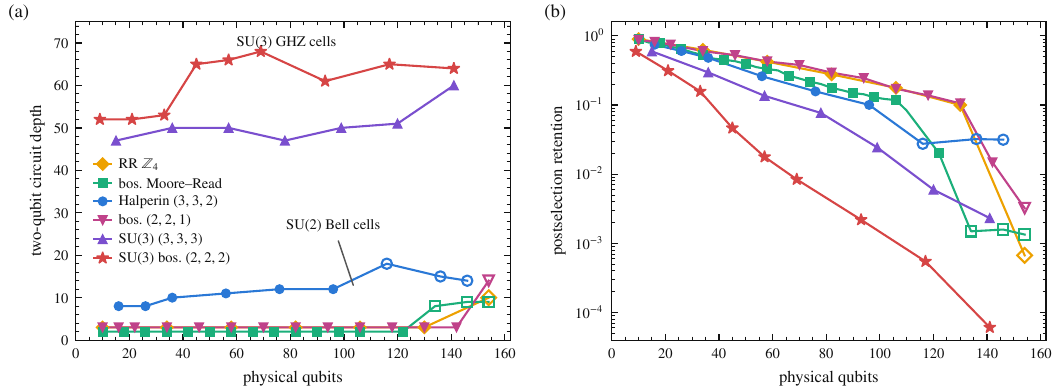}
\caption{The six catalog-extension FQH ladders flown and realized on \texttt{ibm\_kingston}. (a) Transpiled two-qubit depth against qubit count: each family holds a constant algorithmic depth in system size, two for the bosonic Moore--Read, three for the Read--Rezayi $\mathbb{Z}_4$ and the bosonic Halperin $(2,2,1)$, 8 to 12 for the Halperin $(3,3,2)$, and 47 to 68 for the two three-component SU(3) families, whose GHZ cells sit an order of magnitude above the SU(2) Bell cells yet stay flat. Heavy-hex routing raises the depth only at the largest registers (open markers). (b) Postselection retention against qubit count for the same six ladders. The bosonic $(2,2,2)$ curve's deepest register at 141 qubits, coverage-limited on any single pass, completes clean once the passes of both devices are pooled.}
\label{fig:newfam}
\end{figure*}

\subsection{The remaining FQH catalog families}
\label{sec:tail}

We now verify the remaining families of the 18-member catalog (Table~\ref{tab:factory}) in their exact-diagonalization windows.
The Read--Rezayi $\mathbb{Z}_3$ circuits that anchor the scaling ladder of Sec.~\ref{sec:scale} reach retentions of $93$, $80$, and $72\%$ at 8, 13, and 18 qubits, with densities within $0.012$ of exact diagonalization, against $46\%$ retention for the Laughlin circuit at the same electron number.
The Gaffnian behaves equally well at these sizes.
The bosonic $\nu=1/2$ Laughlin state, encoded at two qubits per orbital because the leading $101 \to 020$ squeeze that generates the state requires occupation two, returned densities within $0.04$ of exact diagonalization at all three sizes.
The entangled root and multiplexed junction gates of the Halperin 331 state make its full circuit the most expensive member of the catalog, quantitative at 12 qubits and marginal at 20. Its Bell-cell root family alone, with the exponentially small junction squeeze dropped, scales at two-qubit depth three to 156 qubits (Table~\ref{tab:factory}).
The generalized $\nu=1/4$ Pfaffian requires two degenerate $m=9$ three-body pseudopotential channels (the relative-momentum-9 space is two-dimensional), and its kernel counting follows a two-scale spacing rule. 

The Gaffnian and the Haffnian enter as non-unitary boundary representatives, expected gapless in the thermodynamic limit~\cite{SimonGaffnian}, stressing the counting and circuit machinery beyond the unitary catalog. The Haffnian's non-unitary parent counting is pathological above its natural flux, while its thin-torus representative runs at 11 CZs and $77\%$ retention.

\begin{figure*}[t]
\includegraphics[width=\textwidth]{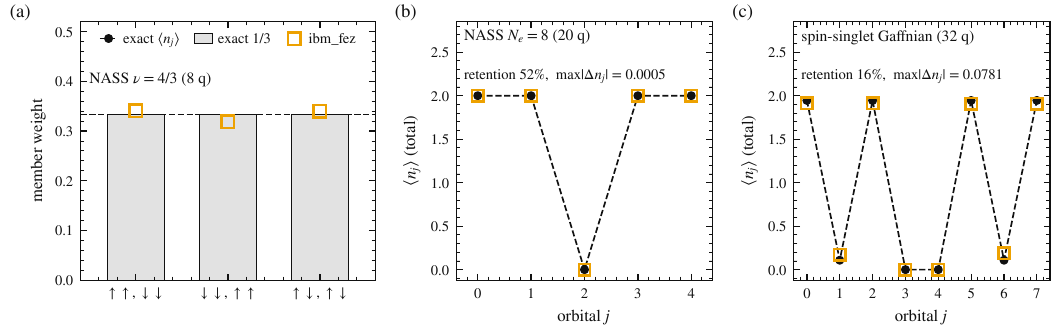}
\caption{Catalog verification of the spinful-boson FQH families on \texttt{ibm\_fez}, 8192 shots per circuit with XY4 dynamical decoupling and Pauli twirling, $(N_\uparrow, N_\downarrow, K)$-postselected. (a) The eight-qubit NASS $\nu = 4/3$ circuit prepares the exact zero mode. The postselected shots put zero weight outside the three-member spin multiplet and reproduce the equal member weights of $1/3$. (b, c) Total densities of the 20-qubit NASS root and the 32-qubit spin-singlet Gaffnian against the exact zero modes (retentions 52\% and 16\%).}
\label{fig:batch4}
\end{figure*}

The two spinful-boson members, the non-Abelian spin-singlet (NASS) state at $\nu = 4/3$ and the spin-singlet Gaffnian, carry entangled spin-triplet multiplets per charge cell and bosonic occupancies up to two.
Their circuits use a dual-rail two-bit-per-mode encoding in which every single bit flip is heralded by particle number alone.
Because the NASS symmetry block contains exactly one state per total spin, the eight-qubit, 20-CZ circuit prepares the parent Hamiltonian's zero mode exactly at every $L_y$.
On hardware, it returned densities within statistical noise of exact ($\max_j |\Delta n_j| < 10^{-4}$) at $79\%$ retention, and its root-family circuit carries the same state to 92 qubits with all 16 charge peaks recovered on the surviving shots.
The 16 squeezing channels of its junction layer collapse to one closed form, the SU(2)-scalar pair hop $-\sqrt{n_L n_R}\, (\sqrt{2})^{[s_L = s_R]} e^{-\kappa^2}/2$.
The spin-singlet Gaffnian, verified at 12 and 32 qubits, needs 154 CZs at $16\%$ retention for the 32-qubit preparation, with densities within $0.08$ of exact, the most gate-intensive of the spinful-boson circuits.
Figure~\ref{fig:batch4} collects these verifications.

Two further clustered members extend the series. The first is the parafermionic Read--Rezayi $\mathbb{Z}_4$ state, one cluster step beyond $\mathbb{Z}_3$ ($\nu=2/3$, five-body pseudopotentials (channel polynomials in Table~S2 of the Appendix~\cite{SM}), root $111100$, $(4,6)$-admissible). Its junction circuit holds two-qubit depth three at 3, 6, and 9 CZ on 10, 16, and 22 qubits with signed overlaps at or above $0.9993$ against the exact zero mode. The second is the bosonic Moore--Read state ($\nu=1$, three-body contact, $(2,2)$-admissible), whose dual-rail-encoded parallel circuit reaches overlap $0.999$ at depth two.
Both pass all four acceptance tests, including an independent cross-check in a second code base (Sec.~S2 of the Appendix~\cite{SM}).

On \texttt{ibm\_kingston}, the $\mathbb{Z}_4$ ladder holds two-qubit depth three from 10 to 130 qubits, with retention falling from $0.899$ to $0.0998$ and the $111100$ root recovered at every electron on every register, from 8 electrons at 10 qubits to 88 at 130. Shots pooled over 10 preparations carry the ladder to a 154-qubit, 104-electron register at routed depth six, with all 104 electrons recovered clean on 363 postselected shots, the largest fractional quantum Hall orbital wavefunction sampled on a processor.
The bosonic Moore--Read ladder holds depth two with zero routing overhead across 26 registers, from 10 qubits at retention $0.892$ to 110 qubits at $0.083$, with the root recovered at every size. Retranspilation extends the depth-two edge to 122 qubits (all 31 peaks at retention $0.020$), and a routed 154-qubit register at depth nine resolves all 39 root orbitals clean on 200 postselected shots.
Constant depth thus persists one cluster size higher, at $k=4$.

The Abelian side of the clustered branch adds two further members: the fermionic Halperin $(3,3,2)$ bilayer at $\nu=2/5$ and the bosonic spin-singlet Halperin $(2,2,1)$ at $\nu=2/3$, both clustered yet carrying only Abelian anyons.
Their kernel-counting identities close exactly, with no excess, against the two-component divisibility formula, at eight and nine flux points. The $(2,2,1)$ count also matches the $k=1$ member of the NASS series.

The two circuits expose different cell anatomies (sketched in Fig.~S1 of the Appendix~\cite{SM}).  
The $(3,3,2)$ root cell carries an independent three-level superposition with the closed-form weight ratio $-2\,e^{-4\pi^2/L_y^2}$. Because the cells stay independent, the hardware ladder runs from 16 to 146 qubits. Through 96 qubits, the two-qubit depth stays 8 to 12, set by the cell anatomy rather than the system size, at retention $0.745$ to $0.101$, with all $2N$ root peaks recovered at every size. The 116-qubit register at routed depth 18 recovers all 24 peaks at retention $0.028$, and two screened \texttt{ibm\_boston} registers extend the family to 136 and 146 qubits at routed depths 15 and 14, recovering all 28 and 30 root peaks at retention $0.032$.
The $(2,2,1)$ cell is a dual-rail Bell block, and its ladder holds two-qubit depth three from 10 to 142 qubits ($N = 2$ to $24$) at retention $0.873$ to $0.0148$. Every root peak is recovered at every size, with a root-family weight of $0.95$ at 142 qubits. A 154-qubit register at routed depth 13 recovers all 52 peaks on 276 postselected shots pooled across two preparations.

Two members extend the catalog to three components: the fermionic Halperin $(3,3,3\,|\,2,2,2)$ at $\nu=3/7$ and the bosonic spin-singlet $(2,2,2\,|\,1,1,1)$ at $\nu=3/4$.
Each root is a six-configuration color-permutation cell, and the two families realize the two one-dimensional irreps of the color group $S_3$.
The fermionic root is the symmetric multiplet, with all six members at the same positive weight, while the bosonic root is the antisymmetric multiplet, the member weight tracking the parity of the color permutation.
Their kernel-counting identities close exactly against the three-color product formula at eight and nine flux points, with no excess even at the per-color flux where a symmetric-power rule would only bound the count. The bosonic $(2,2,2)$ is thus a three-component Abelian state, distinct from the two-component non-Abelian NASS state.

On quantum hardware, the fermionic ladder runs from 15 to 141 qubits ($N=1$ to $7$) at a flat two-qubit depth of 47 to 60 while the CZ count grows linearly from 62 to 480. It stays clean to the full chip: every per-color root peak is recovered at every size, the root-family weight falling only from $1.00$ to $0.87$ at 141 qubits and retention from $0.596$ to $0.0023$.
The bosonic ladder runs from 9 to 141 qubits ($N=1$ to $12$) at a flat depth of 52 to 68 while the CZ count grows from 71 to 971.
Its antisymmetric cell is recovered cleanly to 93 qubits at root-family weight $0.69$ on \texttt{ibm\_kingston}. Two \texttt{ibm\_kingston} passes at 117 qubits kept only 18 and 16 surviving shots, but the device swap onto \texttt{ibm\_aachen} completes the register clean, with all 30 per-color root peaks recovered on 245 postselected shots at root-family weight $0.58$. At 141 qubits, only two \texttt{ibm\_kingston} shots survive and the root is lost, but screened relayouts that exclude the broken readout qubits recover all 36 per-color root peaks in every pass, and pooled coverage completes the register clean. Five passes across three screened layouts on \texttt{ibm\_aachen} and \texttt{ibm\_boston} reach 134 surviving shots, with the pooled on-root density separated from every off-root orbital by a margin of $0.94$.
Figure~\ref{fig:newfam} collects all six extension ladders.  

\section{Quantum hardware measurements of topological responses}

\begin{figure*}[t]
\includegraphics[width=0.8\textwidth]{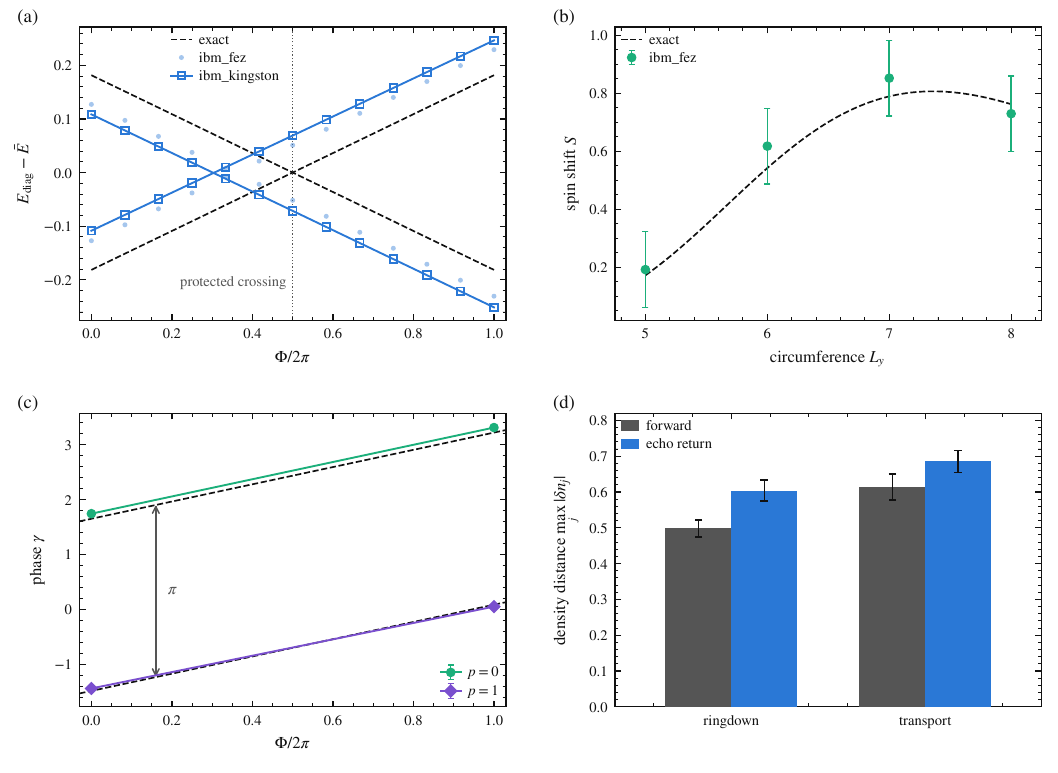}
\caption{Measured topological responses of the prepared states.
(a)~Charge-pump spectral flow of the $\nu=1/3$ Laughlin state on \texttt{ibm\_kingston} (open squares) with the \texttt{ibm\_fez} replication (light points): the diagonal energies of the two adiabatic branches run against threaded flux at measured slopes $+0.3549(11)$ and $-0.3598(10)$ per flux quantum against the exact $\pm0.36$ (dashed), and the pump staircase steps by $0.794(4)$ of the quantized unit charge across the $K$-protected crossing marked at half flux.
(b)~The spin shift of the $\nu=1/2$ Moore--Read state on \texttt{ibm\_fez}, the first geometric response measured on a Moore--Read state: the finite-cylinder $S(L_y)$ tracks the exact truncated-model curve within one standard deviation at all four circumferences, including its maximum.
(c)~Non-Abelian braiding of the Moore--Read $e/4$ quasiholes on \texttt{ibm\_kingston}: the loop-reversal-averaged interference phase of both fusion parities follows the exact slope-$1/4$ reference lines and holds the two channels an Ising monodromy $\pi$ apart, $3.10(3)$ and $3.02(3)$ radians at the flux endpoints, read from the structurally unbiased width-scan re-decode, which runs on \texttt{ibm\_kingston} alone.
(d)~Echo control on \texttt{ibm\_kingston} for the $\nu=1/3$ Laughlin ground state (ringdown) and the 17-orbital $e/3$ wall state (transport): the density distance from the initial profile after the echo, $0.60(3)$ for the ringdown and $0.69(3)$ for transport, is at least as large as the forward move itself, $0.50(2)$ and $0.61(4)$, so the profile motion at roughly 800 CZs per Trotter step is noise dominated.
}
\label{fig:responses}
\end{figure*}

We next measure several topological responses of the prepared states.

\subsection{Charge pumping and spectral flow}

Threading flux through the cylinder pumps the orbitals against fixed confining walls [Fig.~\ref{fig:responses}(a)].
On the lattice, the two adiabatic branches of the $\nu=1/3$ Laughlin state pump are flux-independent squeezed states in dipole sectors that differ by the pumped charge, and they cross exactly at half flux under the protection of $K$ conservation.
Both branch states are 20-CZ circuits, and the entire 13-point spectral-flow curve follows from classically reweighting the two measured configuration distributions against the flux-dependent diagonal Hamiltonian.
The measured charge-transfer slopes are $+0.3549(11)$ and $-0.3598(10)$ per flux quantum against the exact $\pm0.36$, one branch exact within errors and the other $1.4\%$ low. The reference $\pm0.36$ is the finite-size value of the truncated two-branch model on this register rather than the thermodynamic $\nu=1/3$, which it approaches as the droplet grows (Sec.~S6 of the Appendix~\cite{SM}).
The pump staircase steps by $0.794(4)$ of the quantized unit charge, with the interpolated crossing at $\Phi/2\pi = 0.30(10)$, within two standard deviations of the protected half-flux crossing (Sec.~S6 of the Appendix~\cite{SM}).
A replication on \texttt{ibm\_fez} returns the same slopes, $+0.356$ and $-0.358$ per flux quantum, and a crossing at $\Phi/2\pi = 0.36(9)$ that agrees with \texttt{ibm\_kingston} within errors, so the picture replicates across the two devices.

\subsection{Chiral momentum polarization}

From the same shots, we read another topological response, the momentum polarization~\cite{TuZhangQi,ZaletelMongPollmann}, which probes the chirality of the state directly.
A magnetic translation of half the cylinder around its circumference is diagonal in the Landau-orbital basis (orbital $j$ is a momentum eigenstate), so the partial-translation expectation $\lambda(L_y) = \langle \exp[\,i(2\pi/L_y)\sum_{j<j_c} (j - j_c)\, n_j\,] \rangle$ is a linear functional of the occupation-basis populations we already sample.
It costs no basis rotation and no new circuit. We verified it equals the reduced-density-matrix trace to machine precision on the exact Laughlin, Moore--Read, and Read--Rezayi zero modes, and re-analyzed the archived Laughlin counts.
Across three sizes, $N_e = 6$, $8$, and $10$ (16, 22, and 28 orbitals) and on both devices, the measured $\lambda(L_y)$ tracks the exact value to within about $2\%$ in magnitude and $0.03$ radians in phase, at the same postselection retention as the other observables (Fig.~\ref{fig:mompol}).
The observable and the central-charge obstruction are derived in Sec.~S6 of the Appendix~\cite{SM}.

\begin{figure}[t]
\includegraphics[width=\columnwidth]{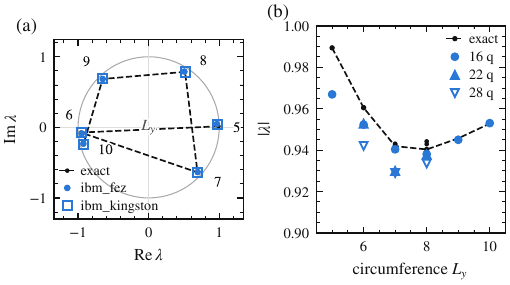}
\caption{Chiral momentum polarization, re-analyzed from the archived Laughlin counts (8192 shots per circuit, $(N,K)$-postselected) at no additional quantum circuit cost. (a) $\lambda(L_y)$ winds around the complex plane as the circumference grows. \texttt{ibm\_fez} and \texttt{ibm\_kingston} track the exact values ($N_e = 6$, 16 qubits). (b) The magnitude $|\lambda(L_y)|$ tracks the exact curve across three circuit sizes, 16, 22, and 28 qubits. The nonzero, $L_y$-dependent phase is the chiral fingerprint: it reverses sign under momentum inversion, and a time-reversal-invariant state would give a real $\lambda$. }
\label{fig:mompol}
\end{figure}
The polarization is chiral: its phase is nonzero and reverses under momentum inversion, whereas a time-reversal-invariant state would return a real value.

\subsection{The Moore--Read spin shift}

The Hall viscosity protocol~\cite{Kirmani2025} [Fig.~\ref{fig:responses}(b)] deforms the cylinder metric and reads the Berry curvature of the ground-state family from finite-difference overlaps measured by an ancilla Hadamard test.
The metric enters the thin-torus state only through a complex squeezing amplitude, so after the gauge-independent reduction the overlap circuits act nontrivially only on the flag register.
For the Moore--Read FQH state, the disjoint blocks reduce each overlap to a product of single-qubit interferences, and the full measurement runs at four two-qubit gates per circuit.
The measured finite-cylinder spin shift, the first geometric-response measurement of a Moore--Read state, tracks the exact truncated-model curve within one standard deviation at all four $L_y$ points, including its maximum, with 96--97\% of shots surviving the working-register check.
The Laughlin replication track of the same job, whose chained flag register costs 94 CZs per circuit, returns a signal damped to half the ideal and quantifies what the recursion structure costs in an interferometric setting.

\subsection{Braiding: the $e/4$ slope and the Ising splitting}

The braiding protocol~\cite{Kirmani2023} [Fig.~\ref{fig:responses}(c)] generalizes from Abelian to non-Abelian quasiholes: two branches of a pinned-wall superposition are prepared, one branch is transported around the cylinder by stepping the pinning potentials, and the accumulated Berry phase is read from quadrature interference circuits. The protocol applies to any catalog state whose quasihole walls the pinning potentials can hold and step.
It is an interferometer in the parent Hamiltonian's quasihole subspace, compiled onto the wall-position register that the pinning potentials define.
A movable $e/4$ wall is a coherent Gaussian superposition over discrete wall positions, and one full transport loop around the cylinder circumference imprints the Berry phase $\gamma = -2\pi\langle q \rangle$.
Dipole quantization forces a structure with no Laughlin analog: moving a wall by one lattice cell necessarily flips the fusion parity of its pair, so fixed-parity walls live on a four-orbital ladder and the two fusion channels sit half a ladder unit apart.
The flux derivative of $\gamma$ therefore measures the quasihole charge, and the parity splitting of $\gamma$ measures the Ising monodromy $\pi$.

We integrate the projector-Hamiltonian adiabatic segments classically and verify that four segments on a three-qubit register reproduce both signatures to $10^{-3}$.
We then promote the segment unitaries to an ancilla-controlled interference circuit of 349 CZs on four qubits, with every circuit checked against the exact overlap to $10^{-9}$ before submission.
On \texttt{ibm\_kingston}, the circuit depth damps the interference contrast to $|\langle \psi_0 | U_{\mathrm{braid}} | \psi_0 \rangle| \approx 0.12$, exactly as forecast, but the damping leaves the measured phases intact.

The raw flux slopes are $0.2344(99)$ on \texttt{ibm\_kingston} and $0.2544(124)$ on \texttt{ibm\_fez}. Recalibrating each phase by its measured quadrature contrast moves both to $0.2500(17)$ and $0.2503(12)$ against the exact $1/4$. The width scan below gives a calibration-free statement.

The fusion-channel splittings read $\{3.20, 2.95, 2.76\}$ and $\{3.11, 2.89, 2.71\}$ radians against the Ising value $\pi$, with per-point errors of $0.08$ to $0.12$ radians, and share a flux-dependent drift on both devices, $-0.44(12)$ on \texttt{ibm\_kingston} and $-0.41(16)$ on \texttt{ibm\_fez}.
The classical reference curve is flat in the finite difference, which rules out the discretization as the source.
A dedicated width-scan systematics run pins the drift to the decode contrast rather than the transported state, removing two thirds of it (Sec.~S8 of the Appendix~\cite{SM}), and loop-reversal averaging, with forward and conjugated paths ideally giving $\pm\gamma$, cancels half of what is left, leaving fusion splittings of $3.10(3)$ and $3.02(3)$ at the flux endpoints.

The flux slopes sit within $0.01$ to $0.03$ of the exact $1/4$ at every width and parity with no recalibration at all.
Both braiding signatures replicate across the two devices and survive the width-scan systematics.

\subsection{Real-time dynamics}

A Trotter step of the truncated Hamiltonian compiles from the same gate set as the variational layers of Sec.~\ref{sec:wall} [Fig.~\ref{fig:responses}(d)].
We quenched the $\nu=1/3$ Laughlin $L_y=5$ ground state and the 17-orbital $e/3$ wall state (the ringdown and the transport quench, respectively) under $H(L_y=7)$ for one and two symmetric Trotter steps, between 800 and 1700 CZs.  
At these depths, the postselected shot fraction falls below $1\%$, and the measured density changes track the exact evolution with spatial correlation $0.88$ to $0.98$ at every evolved point.
Coherent dynamics requires more than that correlation, because the quench thermalizes the densities toward the sector mean, and in-sector depolarization drives them in nearly the same direction (null correlation $0.90$ to $0.98$).
Projecting out the depolarizing direction leaves a residual that agrees with the exact evolution at correlation $\approx 0.4$ at three of the four evolved points, and at the first transport step the agreement is significant: $+0.45$, with the $95\%$ interval $[0.09, 0.62]$ excluding zero on the 91 surviving shots. The remaining points stay suggestive at the surviving shot count.

The experiment that bounds the noise contribution is an echo control: the same step followed by its exact inverse, with a barrier blocking transpiler cancellation, at 1545 and 1657 CZs (the ringdown and transport registers compile to 798 and 826 CZ per step, so their echoes differ).  
The density distance from the initial profile after the echo is at least as large as the forward move itself, $0.60(3)$ against $0.50(2)$ for the ringdown and $0.69(3)$ against $0.61(4)$ for transport, at retention below $1\%$.
The echo control shows that at roughly 800 CZs per Trotter step the profile motion is noise dominated, apart from that one significant residual.

\section{The route to the isotropic point}
\label{sec:wall}

Apart from deliberate excursions, all the results above operate in the thin-torus window where the representatives are faithful.
The isotropic scenario is a different regime, and we describe its measured distinction quantitatively here. We reach it by two routes: a Hamiltonian variational circuit, and the thin-torus construction itself with its bond dimension raised one qubit at a time. The key difference from the thin-torus circuits is cost, gate count against device noise rather than structure.
A five-parameter Hamiltonian variational circuit prepares the $L_y = 10$ Laughlin state at the same size and operating point as the trapped-ion experiment~\cite{Shen2026}.
We compress it from 1150 to 697 CZs at fidelity $0.919$ by eliding gates whose inputs are exactly known and by evicting the long-range scattering layer through fermionic swaps.
On heavy-hex hardware, it survives postselection at $0.8\%$, and the postselected string order reads $+0.119$ against the exact $+0.374$ and the depolarized baseline $+0.012$.
The signal is unambiguous but heavily damped.

The two routes have different failure modes at $L_y = 10$.
The error of the fixed thin-torus circuit is systematic, the representative's own bias, with an overlap near $0.80$ against the true ground state.
Its measured string order $+0.49$ matches its own target to better than $0.01$ while overshooting the true value $+0.374$ by $+0.12$.
The error of the variational circuit is stochastic: its ideal fidelity is the higher one, $0.919$, but 697 CZs of device noise damp the measured string order to $+0.119$, an error of $-0.26$.
On present hardware, the thin-torus route is therefore the more accurate one even beyond its faithful window, because the noise penalty of the deep circuit exceeds the representative bias.
This ordering is set by the device error rate rather than by principle, and the bond-dimension interpolation below measures the error rate at which the variational route overtakes.

The same barrier appears in both preparation routes and on both processors, and its origin is structural.
The nonzero chiral central charge of FQH order provably places it beyond any finite-depth local circuit in the thermodynamic limit, unitary or measurement-assisted with feedforward~\cite{ChiralNoGo1,ChiralNoGo2}. The theorems constrain circuits local in the two-dimensional geometry; the tensor-network obstruction below is their orbital-space counterpart. 
The no-go theorems extend to circuits an obstruction long known for tensor networks: chiral free-fermion states admit no gapped local parent Hamiltonian~\cite{WahlChiralPEPS,DubailRead}.
As the circumference grows, each exact preparation route meets its own asymptotic obstruction: the sparse-state route~\cite{Wu2026} through the growth of the wavefunction's support, the variational route through fidelity collapse, and any exact tensor-network route through a bond dimension that rises exponentially with circumference~\cite{ZaletelMong}.
The barrier concerns only the strict thermodynamic limit. The isotropic bulk itself is already reached at a circumference of 10 to 12 magnetic lengths, where the exact matrix-product bond dimension of the $\nu=1/3$ state is still single-digit.

Our thin-torus circuits (the construction applies catalog-wide; the hardware demonstration below is for $\nu=1/3$) are the leading, bond-dimension-two member of exactly that matrix-product family, and adding one bond qubit at a time carries the same construction from the quasi-one-dimensional window toward the isotropic crossover at a cost linear in particle number, up to the asymptote set by the no-go theorems.
We verified this construction classically: at bond dimension two, the circuit reproduces the thin-torus preparation exactly, and raising the bond dimension to eight reaches the isotropic bulk at $L_y = 10$ with infidelity below $10^{-4}$, at a gate count that grows linearly in $N$ and sits 7 to 48 times below the exact sparse-state route at $N_e = 5$ and $6$ (Sec.~S9 of the Appendix~\cite{SM}).
The remaining barrier is gate count against device noise rather than bond dimension, which for $\nu = 1/3$ stays modest until well past the isotropic crossover.
We tested this bond-dimension knob on \texttt{ibm\_marrakesh} at $L_y = 10$, past the thin-torus window [Fig.~S6 of the Appendix~\cite{SM}]: the postselected density's distance to the exact isotropic state falls from $0.096$ at $\chi = 2$ to $0.049$ at $\chi = 4$.
The $\chi = 8$ circuit, at 1568 CZs, is noise dominated.

\section{Discussion}

In summary, we have put forth a general approach to catalog and construct FQH states at unprecedented scale and variety. Together, the Hamiltonian catalog, the closed-form thin-torus circuits, and symmetry-verified sampling provide a common workflow across a broad range of FQH families. Our construction yields two structural results: closed-form squeezing amplitudes and a clustering-based depth dichotomy. In particular, most of the more sophisticated FQH states with exotic clustering properties, whose experimental or even numerical demonstration has proved elusive so far, are prepared with constant circuit depth at scale. All implementations reported here lie in the thin-torus regime. One added bond qubit carries the same registers beyond it, halving the hardware distance to the exact isotropic state, at a cost set by the chiral order itself (Sec.~\ref{sec:wall}).  

Symmetry-selected occupation-basis measurements provide access to four responses: fractional charge, hidden order, spectral flow, and chiral momentum polarization.
For the clustered states, the $e/4$ and $e/5$ charges are exact in every retained shot through the manifold-invariance mechanism, while dipole quantization ties the wall position to its fusion parity.
Tables~\ref{tab:factory} and~\ref{tab:responses} summarize the preparations and measurements.
The 154-qubit Read--Rezayi $\mathbb{Z}_4$ implementation demonstrates the accessible register scale, while the Moore--Read interferometry demonstrates the range of non-Abelian observables.

The interferometric protocols extend to two-pair geometries and four-anyon fusion-space tomography, where the Ising even-odd effect becomes accessible at the few-hundred-CZ scale of the one-pair interferometer measured here~\cite{SternHalperin2006,BondersonKitaevShtengel2006}.
The same parent-Hamiltonian kernels can also seed exact sparse-state circuits away from the thin-torus regime.
A particularly direct next step is a local coherence witness on a clustered ladder, requiring only a constant number of measurement settings to distinguish the coherent squeezed state from an incoherent mixture with the same root density.

At fixed circumference, the circuit depth of an FQH trial state is controlled by the factorization structure of its thin-torus root.
Clustered states admit disjoint local preparation blocks, whereas the Laughlin roots generate a sequential chain.
The parent Hamiltonian makes this distinction operational by supplying the circuit amplitudes, symmetry checks, quasihole sectors, and response protocols within a single framework.
Quantum processors can therefore access a broad catalog of Abelian and non-Abelian FQH representatives at low logical depth, using full-state circuits for the disjoint families and root-family circuits for the entangled-cell ones. The crossover toward the isotropic liquid then exposes the additional entanglement and gate resources required by chiral topological order.

\section*{Acknowledgments}
C.X. and Y.Z. were supported by the Max Planck partner lab
for quantum materials. CHL acknowledges support from the Ministry of Education, Singapore (Awards No.~MOE-T2EP50224-0007 and MOE-T2EP50224-0021).
The authors acknowledge the use of IBM Quantum services for this work. The views expressed are those of the authors, and do not reflect the official policy or position of IBM or the IBM Quantum team.

\section*{Data availability}
All circuits, raw and decoded counts, job identifiers, calibration snapshots, and analysis scripts are archived with the project repository, including a registry of every hardware job on the IBM Heron devices \texttt{ibm\_fez}, \texttt{ibm\_kingston}, \texttt{ibm\_marrakesh}, \texttt{ibm\_aachen}, and \texttt{ibm\_boston}.

\bibliography{ref}

\end{document}